\documentclass[a4paper,fleqn,usenatbib]{mnras}

\usepackage{newtxtext,newtxmath}
\usepackage{chngcntr}

\usepackage[T1]{fontenc}
\usepackage{ae,aecompl}

\usepackage{graphicx}	
\usepackage{amsmath}	
\usepackage{amssymb}	

\title[Structure in the Disk of epsilon Aurigae]{Structure in the Disk of epsilon Aurigae -- Analysis of ARCES and TripleSpec spectra from the 2010 eclipse}

\author[J. Gibson, et al.]{
Justus L. Gibson$^{1}$\thanks{E-mail: justus.gibson@du.edu},
Robert E. Stencel$^{1}$,
William Ketzeback$^{2}$, 
John Barentine$^{3}$,
\newauthor
Jeffrey Coughlin$^{4}$,  
Robin Leadbeater$^{5}$, and
Gabrelle Saurage$^{6}$
\\
$^{1}$Department of Physics and Astronomy, University of Denver, 2112 E Wesley Ave, Denver, CO 80208, USA\\
$^{2}$Apache Point Observatory, 2001 Apache Point Rd, Sunspot, NM 88349, USA\\
$^{3}$International Dark Sky Association, 3223 N 1st Ave, Tucson, AZ 85719, USA\\
$^{4}$SETI Institute, 189 Bernardo Ave, Suite 200, Mountain View, CA 94043, USA\\
$^{5}$Three Hills Observatory, UK\\
$^{6}$SOFIA/USRA\\
}

\date{Accepted XXX. Revised 1 June 2018; in original form 23 March 2018}
\pubyear{2018}

\begin{document}
\label{firstpage}
\pagerange{\pageref{firstpage}--\pageref{lastpage}}
\maketitle

\begin{abstract}
Worldwide interest in the recent eclipse of epsilon Aurigae resulted in the generation of several extensive data sets, including high resolution spectroscopic monitoring.  This lead to the discovery, among other things, of the existence of a mass transfer stream, seen notably during third contact. We explored spectroscopic facets of the mass transfer stream during third contact, using high resolution spectra obtained with the ARCES and TripleSpec instruments at Apache Point Observatory. One hundred and sixteen epochs of data were obtained  between 2009 and 2012, and equivalent widths and line velocities measured for high versus low eccentricity accretion disk lines. These datasets also enable greater detail to be measured of the mid-eclipse enhancement of the He I 10830{\AA} line, and the discovery of the P Cygni shape of the Pa-$\beta$ line at third contact.  We found evidence of higher speed material, associated with the mass transfer stream, persisting between third and fourth eclipse contacts.  We visualized the disk and stream interaction using SHAPE software, and used CLOUDY software to estimate that the source of the enhanced He I 10830A absorption arises from a region with n$_H$ = 10$^{11}$ cm$^{-3}$ and temperature of 20,000 K, consistent with a mid-B type central star. 
Van Rensbergen binary star evolutionary models are somewhat consistent with the current binary parameters for their case of a 9 plus 8 solar mass initial binary, evolving into a 2.3 and 14.11 solar mass end product after 35 Myr.  With these results, it is possible to make predictions which suggest that continued monitoring prior to the next eclipse (2036) will help resolve standing questions about the mass and age of this binary.\\
\end{abstract}




\section{Introduction}
\label{sec:intro}


The exceptional binary star, epsilon Aurigae, exhibits two year-long eclipses every 27.1 years.  The eclipse of the F supergiant primary star is now understood to be caused by an opaque disk hiding a companion star (see \citet{Ste12} and references therein).  An international effort to monitor all aspects of the 2009-2011 eclipse resulted in substantial archives of new data, including those involving high resolution spectroscopic monitoring, e.g. \citet{Str14} and others.  Here we report on sampling of the high resolution spectroscopic archive generated during 2009-2011 eclipse, by the ARCES instrument at Apache Point Observatory \citep{Wan03}.  As discussed by \citet{Str14}, selected spectral lines exhibit eclipse or non-eclipse behaviors, and show differential velocity effects.  Our purpose is to confirm and extend their findings, in light of continuing developments about the nature of this enigmatic binary star and disk system. 

The epsilon Aurigae system is a member of the zeta Aurigae family of long period eclipsing binary stars that feature an evolved supergiant star plus a B dwarf star companion.  That group of stars is part of the still larger group called Algol stars, wherein a higher luminosity, evolved but less massive primary star is eclipsed by a more massive main sequence companion star, indicating that mass transfer has occurred.

Among the developments resulting from study of the 2009-2011 eclipse of epsilon Aurigae was recognition of third contact phenomena, interpreted as a mass transfer stream impacting the disk \citep{Gri13}.  Third contact occurred during spring 2011, circa RJD 55,620 (RJD = Reduced Julian Date, Table 1 and \citet{Ste12}.  Evidence for this stream was seen in low excitation Fe I lines and, curiously, lines of rare earth elements, blueshifted up to 45 km/sec.  Corroborating observations of spectral line radial velocity curves near third contact, showing similar effects, were reported by \citet{Kam13} and \citet{Sad13}. \citet{Str14} presumably would have detected similar spectral features if their robotic STELLA spectrometer had not been off-line during those portions of the eclipse. Among the significant findings from STELLA \citep{Str14} were the differences among so-called high and low eccentricity lines during disk transit -- a point we shall return to later in this paper.  ARCES observers maintained a steady cadence through all of the eclipse phases and thus provide a more detailed picture of third contact phenomena.  For completeness, we note that the well-studied K I 7699$\AA$ and the saturated Na I D lines \citep{Lead12} do not show evidence for specifically third contact phenomena of the type described here. 

In this paper we will discuss the new spectroscopic observations, followed by SHAPE and CLOUDY models for the system (Sections 2 and 3) and then provide an evolutionary context (Section 4) and Conclusions in Section 5.

\section{Observations with APO ARCES and TripleSpec instruments}
The ARCES instrument is an echelle spectrograph commissioned in 1999 for use with the ARC (Astrophysical Research Consortium) 3.5m telescope located at the Apache Point Observatory \footnote{http://www.apo.nmsu.edu/arc35m/Instruments/ARCES/  DOI: 10.1117/12.461447}.  ARCES main optics consist of an off-axis parabolidal collimator, an echelle grating, two cross-dispersing prisms, and an f/2.7 Schmidt camera with achromatic correctors. Since its commissioning, ARCES has been primarily used for observations of stellar abundances and as a tool for surveying diffuse interstellar bands (DIBs). Important features of this instrument include: a resolution of 33,000 (9 km/s) with the 1.6 inch slit, a spectral range of 3200A-10000A, an S/N $\geq$ 3000, and remote operability. In addition, the efficiency of the entire  telescope plus spectrograph system has been determined to be greater than 2.2\% at 647nm and the spectrograph has been estimated to have an efficiency between 8\% and 2\% at 630nm. Some limitations of this instrument include a slight drift throughout nightly observations of about 0.5 pixels over 10 hours. This drift occurs mostly during the first half of the night and is likely due to thermal changes in the double-prism system as temperatures rise and fall. This has been accounted for by interpolating values between hourly calibrations. Steps in the data reduction described by Wang et al. (2003) include: (1) extracting spectra from raw CCD frames; (2) applying deblazing and wavelength standardization; (3) applying heliocentric RV-corrections, but no telluric corrections were attempted. To achieve high-accuracy RV measurements with the echelle spectrograph, a Thorium-Argon (Th-Ar) exposure was automatically obtained after every science exposure as part of the built-in calibration procedure.  The 9 km/s velocity resolution translates to a spectral precision of 0.15{\AA} at 5000{\AA} and 0.20{\AA} at 6500{\AA}. Overall, ARCES has been proven to be a stable, reliable instrument that has many uses in astronomical observations.

Similarly, TripleSpec \citep{Wil04} is a near-IR spectrometer built for the ARC 3.5 meter telescope at Apache Point Observatory \footnote{APO: https://tinyurl.com/yb77xrgj}. Spectral coverage ranges between 0.95 and 2.46 microns, with 2.1 spectral pixel resolution of 3500 with the 1.1 arcsec slit. Triplespec has no internal calibration sources and relies on bright sky lines for wavelength calibrations. The TripleSpec instrument includes a data extraction routine called TripleSpecTool, which is a modified version of the Spextool package developed by Michael Cushing \citep{Cus04}. The package coverts TripleSpec images into one-dimensional calibrated spectra corrected for telluric absorptions using a method developed by \citet{Vac03}. Data reduction proceeds through a number of steps including: (1) Importing raw fits files from TripleSpec images, (2) generating a flat field from observations of Bright Quartz lamps for wavelength calibrations, (3) flux-calibrating, removing intrinsic stellar features, and removing telluric absorption lines. More information about the TripleSpec instrument and the associated data reduction process can be found in the papers cited above. 

ARCES and TripleSpec observations of epsilon Aurigae are listed in Table 1, including date and Reduced Julian Date plus an estimate of the signal to noise ratio (SNR) based on continuum signal statistics as described below.. Observations of epsilon Aurigae were obtained with ARCES beginning in 2009 Feb 16 (RJD 54878, where RJD = JD - 2,400,000) and continuing every few nights, through 2011 Dec 10	(RJD 55905). These span key times during the eclipse cycle, including ingress (2009 Aug 15, RJD 55060), second contact (2010 Feb 15, RJD 55250), mid-eclipse (2010 July 30, RJD 55400), third contact (2011 March 15, RJD 55620) and end of eclipse (2011 Aug 30, RJD 55800). 

\begin{table*}
\centering
\caption{ARCES and TripleSpec observation log for $\epsilon$ Aurigae} 
{\scriptsize
\begin{tabular}{ccccccc}
\hline
{Date of Observation} & {RJD*} & {SNR} & & {Date of Observation} & {RJD*} & {SNR} \\
yearmonthday & & & & yearmonthday & &  \\ \hline
090216 &	54878 &	417 & &	 	101213 & 	55543 & 	 385 \\
090311 &	54901 &	400 & &	 	101216 & 	55547 & 	 370 \\
090408 &	54929 &	400 & &	 	101220 & 	55551 & 	 385 \\
090410 &	54931 &	400 & &	 	101226 & 	55556 & 	 385 \\
090412 &	54933 &	370 & &	 	101229 & 	55559 & 	 385 \\
090413 &	54934 &	370 & &	 	110104 & 	55565 & 	 357 \\
090414 &	54935 &	400 & &	 	110108 & 	55569 & 	 400 \\
090505 &	54956 &	263 & &	 	110114 & 	55575 & 	 345 \\
090729 &	55041 &	200 & &	 	110121 & 	55582 & 	 385 \\
090804 &	55047 i &	238 & &	 	110205 & 	55597 & 	 385 \\
090908 &	55082 &	227 & &	 	110211 & 	55603 & 	 333 \\
090924 &	55098 &	323 & &	 	110214 & 	55606 & 	 385 \\
091002 &	55106 &	333 & &	 	110222 & 	55614 & 	 345 \\
091027 &	55131 &	345 & &	 	110304 & 	55624 iii &  357 \\
091031 &	55135 &	345 & &	 	110307 & 	55627 & 	 270 \\
091105 &	55140 &	345 & &	 	110316 & 	55636 & 	 303 \\
091108 &	55143 &	357 & &	 	110320 & 	55640 & 	 303 \\
091118 &	55153 &	357 & &	 	110329 & 	55649 & 	 313 \\
091126 &	55161 &	333 & &	 	110401 & 	55652 s & 	 323 \\
091207 &	55172 &	370 & &	 	110402 & 	55653 & 	 323 \\
091220 &	55185 &	370 & &	 	110406 & 	55657 & 	 161 \\
091225 &	55190 &	385 & &	 	110413 & 	55664 & 	 323 \\
100101 &	55197 &	385 & &	 	110416 & 	55667 & 	 333 \\
100104 &	55200 &	213 & &	 	110428 & 	55679 & 	 357 \\
100111 &	55207 &	417 & &	 	110503 & 	55684 & 	 345 \\
100117 &	55213 &	357 & &	 	110729 & 	55771 & 	 179 \\
100130 &	55226 &	370 & &	 	110802 & 	55775 & 	 244 \\
100213 &	55240 &	417 & &	 	110805 & 	55778 & 	 128 \\
100225 &	55252 ii &	345 & &	 	110813 & 	55786 & 	 161 \\
100307 &	55262 &	294 & &	 	110823 & 	55796 & 	 217 \\
100308 &	55263 &	286 & &	 	110827 & 	55800 iv & 	 58 \\
100312 &	55267 &	435 & &	 	110831 & 	55804 & 	 217 \\
100322 &	55277 &	400 & &	 	110920 & 	55824 & 	 250 \\
100331 &	55286 &	370 & &	 	110924 & 	55828 & 	 286 \\
100407 &	55293 &	345 & &	 	111007 & 	55841 & 	 286 \\
100410 &	55296 &	417 & &	 	111011 & 	55845 & 	 278 \\
100424 &	55310 &	303 & &	 	111016 & 	55850 & 	 313 \\
100425 &	55311 &	270 & &	 	111023 & 	55857 & 	 286 \\
100427 &	55313 &	303 & &	 	111026 & 	55860 & 	 278 \\
100722 &	55399 mid &	161 & &	 	111030 & 	55864 & 	 313 \\
100811 &	55419 &	196 & &	 	111102 & 	55867 & 	 263 \\
100821 &	55429 &	189 & &	 	111104 & 	55869 & 	 263 \\ 
100905 &	55444 &	294 & &	 	111107 & 	55872 & 	 196 \\
100912 &	55451 &	294 & &	 	111111 & 	55876 & 	 294 \\
100924 &	55463 &	333 & &	 	111116 & 	55881 & 	 286 \\
100929 &	55468 &	323 & &	 	111119 & 	55884 & 	 270 \\
101007 &	55476 &	345 & &	 	111129 & 	55894 & 	 270 \\
101011 &	55480 &	323 & &	 	111210 & 	55905 & 	 345 \\
101015 &	55484 &	345 & &	 	111216 & 	55911 & 	 345 \\
101018 &	55487 &	313 & &	 	111228 & 	55923 & 	 323 \\
101026 &	55495 &	333 & &	 	120101 & 	55927 & 	 125 \\
101029 &	55498 &	385 & &	 	120106 & 	55932 & 	 357 \\
101103 &	55503 &	370 & &	 	120113 & 	55939 & 	 278 \\
101107 &	55507 &	357 & &	 	120131 & 	55957 & 	 357 \\
101108 &	55508 &	385 & &	 	120209 & 	55966 & 	 141 \\
101111 &	55511 &	370 & &	 	120214 & 	55971 & 	 345 \\
101116 &	55516 &	400 & &	 	120305 & 	55991 & 	 345 \\
101120 &	55520 &	357 & &	 	120323 & 	56009 & 	 370 \\
101122 &	55522 &	345 & &	 	120406 & 	56023 & 	 313 \\
101124 &	55524 &	323 & &	 	120409 & 	56026 & 	 167 \\
101204 &	55534 &	182 & &	 	120428 & 	56045 & 	 294 \\
101209 &	55539 &	370 & &	 			 & 			& 		\\
\hline
\end{tabular}
}
\\
*RJD = J.D. -- 2,400,000. i, near first-contact; ii, second-contact; mid, near mid-eclipse; iii, third-contact; s, stream prominent; iv, fourth-contact
\\
\end{table*}  

\begin{table*}
\centering
\caption{Atomic Line Parameters} 
{\scriptsize
\begin{tabular}{ccccc}
\hline
{Species} & {Rest Wavelength, \AA} & {Multiplet, eV lower} & {f-value} & { Reference} \\ 	
\hline
{\bf Uneclipsed lines} 	&	 	& 		& 			&  			\\
Mg II 	& 	4481.130	& 2D 5/2-2Fo 7/2, 8.86 		& 0.935 & SE \\
Mg II   &   4481.327	& 2D 3/2 - 2Fo 5/2, 8.86		 	&  0.981 & SE \\
Fe I     &	5615.6		& z5Fo-e5D, 3.33 	& 0.010 	& 	k \\
Si II 	& 	6347.10 	& 2S-2Po, 8.12		& 	0.705 	&  	s, k \\ 
Si II	&	6371.36	& 2S-2Po, 8.12		&	0.414	&	s \\
Fe II	&   6417.46		& 3d6-y2Fo, 8.63	&	--         &  -- \\
Fe II	&	6432.68	& 3d6 - 4P, 10.93	&  0.015	&	--  \\
\hline
{\bf Eclipsed lines} 	& 	&			& 			& 		  \\
Fe I 	&	4132.06	& a3F-y3Fo, 1.61	&	0.042	&	s \\
Fe II	&	4178.86*	& b4P-z4Fo, 2.58	&	6E-4	&	s \\
*blend & Cr I 4179.43 & 					& 			& 		   \\
Fe I 	&	4202.03	& a3F-z3Go, 1.48	&	0.022	&	s \\
Fe II	&	4303.18	& b4P-z4Do, 2.70	&	6E-4	&	s \\
Ti II	&	4468.492	& a2G9/2 - z2Fo7/2, 1.13 & 0.024 & SE \\

Fe II	& 	4555.89	& b4F-z4Fo, 2.82	&	7E-4	&	s \\
Fe II	& 	4629.33	& b4F-z4Fo, 2.81	&	6E-4	&	s \\
Fe II	&	4923.92	& a6S-z6Po, 2.89	& 	0.010 	& 	s,k \\ 
Fe I     &	5110.4		& a1H-z1Ho, 3.57 	& 0.000 	& 	k \\
Fe I  	&	5168.90	& a5D-z7Do, 0.05	&	1.5E-5	&	Fig. 5 \\
Fe II	&	5169.03	& a6S-z6Po, 2.89	&	0.023	&	s \\
Fe I  	&	5169.29	& c3F-t3Do, 4.07	&	-- 		&	s \\
Fe II	&	5169.80	& f4D-2[3]o, 10.50	&	0.008	&	s \\	
Fe I 	&   5615.64	& z5Fo5 - e5D4, 3.33 & 0.10  & 	k \\ 
K I  &  7698.96  & 3p6 4s - 3p6 4p, 0.00 & 0.33 & s \\
\hline
\end{tabular}
}
\\
Data from http://physics.nist.gov/PhysRefData/ASD/lines \\
s = Strassmeier et al. 2014; k = Kambe et al. 2013; SE = Struve and Elvey 1930 
\\
\end{table*}


\section{Data Analysis}
Each spectrum obtained from the ARCES instrument was analyzed using a software tool called VOSpec.  VOSpec was developed by ESA, for use in analyzing spectra from a variety of data providers, said data having varying wavelength and flux units.  VOSpec has many tools built into the program such as line fitting, redshift corrections, as well as the ability to calculate equivalent widths and radial velocities. In order to determine the overall quality of each spectrum, a standard deviation function was applied in VOSpec for the featureless continuum between two Si II lines at 6347{\AA} and 6371{\AA}.  This interval appears relatively free of any telluric or stellar features.  Thus, determining the standard deviation of the continuum gives a measure of the signal-to-noise ratio of each spectrum.  To do this, a range was manually selected in VOSpec ranging from approximately 6350{\AA} to 6368{\AA} and then the standard deviation function was applied to the selected region. The results of these tests can be seen in Table 1, reported as SNR, which is the inverse of the continuum interval standard deviation, assuming the usual statistics.  The majority of the spectra have SNR in excess of 200.  Based on previous work cited above, we have elected to study selected spectral lines in the ARCES and TRIPLESPEC spectra of epsilon Aurigae. Certain atomic properties from the online NIST Atomic Spectra Database are given in Table 2.

\subsection{High excitation lines of Si II and Fe II}
 	VOSpec was used to calculate equivalent widths as well as wavelength of line intensity minima which were, in turn, used to calculate radial velocities of certain Si II, Fe I, and Fe II spectral lines investigated by \citet{Str14}. The equivalent width function of VOSpec works by selecting a wavelength range for a particular line of interest and then applying the equivalent width function. For line velocities, the minimum of each line was manually determined and then a Doppler velocity calculation was performed to deduce the radial velocity. The Si II behavior will be referred to as a \textit{Type 1} variation where we see a gradual decrease in radial velocity from +10km/s to --15km/s over a 1200 day period, giving an deceleration of +0.021 km/s/d. The measured equivalent width of 0.57 +/- 0.03{\AA} suggests that despite the velocity changes, the integrated line formation regions are largely unchanged with time.

\begin{figure} 
\begin{center}
\includegraphics[width=1.0\linewidth] {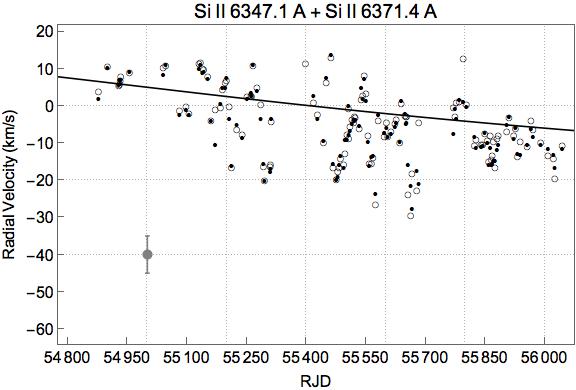}
\caption{Si II 6347.1{\AA} and Si II 6371.4{\AA} velocities  represented by open and closed circles, respectively. The velocity variations are typical of out of eclipse variations associated with non-radial oscillations of the F star photosphere. The error bar in the lower left corner of this figure and subsequent figures indicates a +/- 5 km/s uncertainty. The solid line is the orbital solution from Stefanik et al. 2010. }
\label{fig:siIIrv}
\end{center}
\end{figure}

Velocity variations in Fe II 6417 + 6432{\AA} are similar to Si II (Figure~\ref{fig:siIIrv}) with the exception of a large, negative spike at third contact (RJD 55600), during which a secondary profile component was present. The equivalent widths are essentially unchanged despite the variations in velocity. Additional pairs of high energy lines are the Mg II 4481{\AA} and Fe II 6417/6432{\AA} lines with velocity variations similar to Si II.
    
\begin{figure} 
\begin{center}
\includegraphics[width=1.0\linewidth]{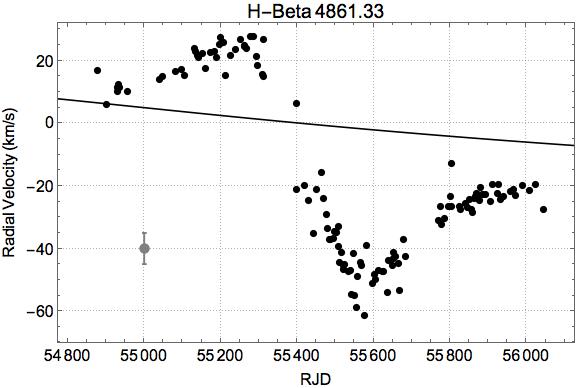}
\caption{Radial velocity of flux minima in H$\beta$, an optically thick line, showing a simple velocity variation interpreted as disk rotation, sampled during eclipse. The solid line is the orbital solution from Stefanik et al. 2010.}
\label{fig:hbrv}
\end{center}
\end{figure}

In contrast, H$\beta$ shows a lot of variation in velocity and equivalent width (Figure~\ref{fig:hbrv}). We interpret this as revealing mostly disk rotation combined with F-star orbital motion. This is a \textit{Type 2} variation characterized by a velocity decrease from 20 km/s to 50 km/s over a period of 400 days around mid-eclipse, giving a deceleration of 0.075 km/sec/day, or 8.68 x 10$^{-4}$ m sec$^{-2}$.

\begin{figure} 
\begin{center}
\includegraphics[width=1.0\linewidth]{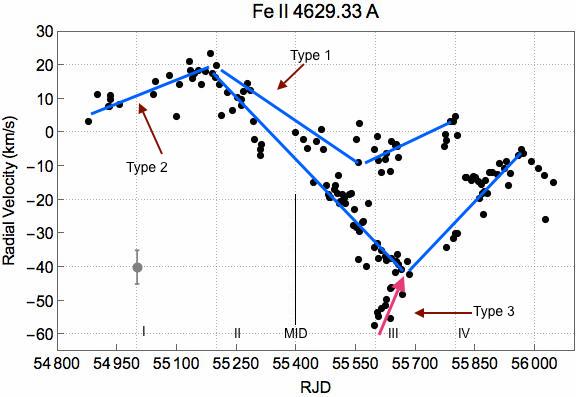}
\caption{Radial velocity of flux minima in Fe II 4629{\AA} .  We see the disk rotation sine-like curve on the right,
as well as another branch, near zero velocity, prominent after mid-eclipse, and we also observe a third contact high velocity component, prominent at RJD 55600. This is a \textit{Type 3} variation characterized by the the Type 2 sine variation plus an extra infall stream at RJD 55600-55700 (acceleration = 0.27 km/s/day, or 3.13 x 10$^{-3}$ m sec$^{-2}$ with opposite sign and ten times larger than the H-$\beta$ value shown in Figure~\ref{fig:hbrv}). The lines represent the Type 1, 2, and 3 velocity variations discussed previously. See text for details}
\label{fig:4629interp}
\end{center}
\end{figure}

Figure~\ref{fig:4629interp} shows the velocity variations observed in the Fe II 4629.33A line, revealing a complex interplay of several dynamically significant features in the disk.  By measuring the progression of each feature appearing and associated with this line (and others), we can trace the following three features: (1) a lower amplitude, sinusoidal variation during eclipse - redshifted to $\sim$15 km/sec near second contact, then blueshifted to $\sim$--10 km/sec near third contact; (2) a higher amplitude variation, initially indistinguishable from (1) until after second contact, but then reaching $\sim$-40 km/sec near third contact, and (3) a dramatic but short lived extreme blue-shifted feature only at third contact, appearing close to -60 km/sec but merging with feature (2) soon thereafter. 

We interpret these three features as follows: the upper curve represents the F star orbital motion (amplitude a few km/sec), while the middle curve represents rotation of a disturbed portion of the disk (amplitude tens of km/sec), perhaps related to the extra feature near RJD 55600 - the impact of a mass transfer stream, briefly illuminated by the background F star during third contact.  This third feature is the mass transfer stream, discovered by \citet{Gri13} but not revealed with this degree of detail, prior to ARCES coverage.  Velocity curves presented by \citet{Str14} drew our attention to these high velocity features, but their temporal coverage was limited by instrument downtime, close to third contact.  We posit that the low and high eccentricity lines, as described by \citet{Str14}, are defined by variations of type 1 and type 2, respectively, in our nomenclature.  Variation of type 3 is the disrupted portion of the disk, caused by stream impact primarily near azimuths around third contact. 

\subsection{Velocity Changes of Selected Disk/Stream Sensitive Lines}

\citet{Sad13} identified a series of lines that showed clear changes due to the passage of the disk.  Their radial velocity plots show several lines, identified in Table 2 and discussed below, with extreme blueshifts during third contact times, but with late eclipse coverage gaps in their robotic series.  We report two kinds of velocity variation here: those showing disk motion, and those showing disk plus stream motions.  First, there are the disk-only set of lines, typically having the lowest excitation potentials (less than 2 eV) and not showing velocities much in excess of --40 km/sec, for example H$\beta$ in Figure 2. The hydrogen lines are optically thick and sensitive to larger scale phenomena in and around the disk and the F star wind \citep{Mou12}. The second group of lines show the disk motion, plus evidence for an infall stream, close to third contact (RJD 55,600), with velocities around -60 km/sec (see Figure~\ref{fig:4629interp}).

\subsection{Line Profile Evolution of Selected Lines}

\begin{figure*} 
\centering
\includegraphics[width=.5\textwidth]{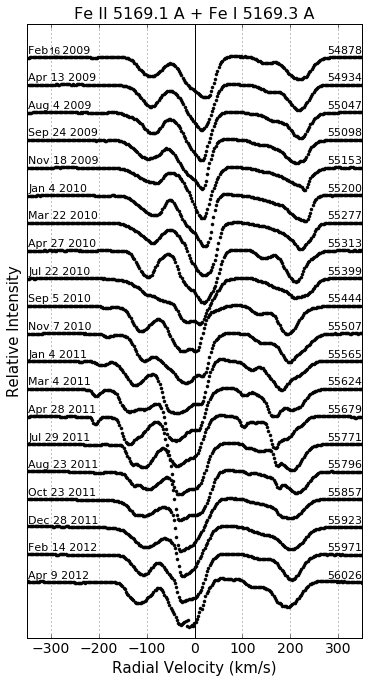}\hfill
\includegraphics[width=.5\textwidth]{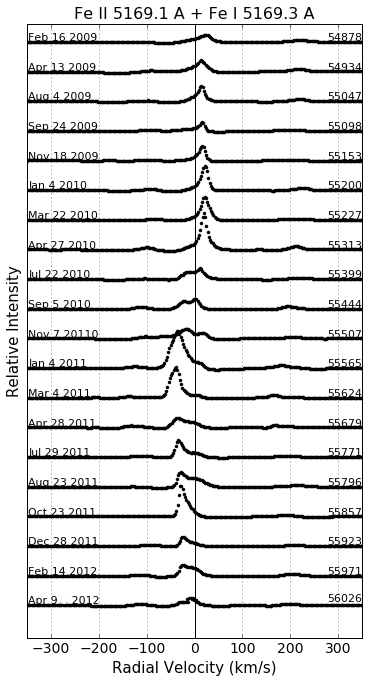}\hfill
\caption{The figure on the left shows a multiplot of Fe II 5169{\AA}  with dates spanning the whole period of monitoring of the eclipse.  As can be seen, Doppler redshift becomes prominent around second contact (RJD 55200) whereas near mid-eclipse (RJD 55480) we begin to see line-doubling and blue-shifting of the central line. The right-most figure shows the results of a first-order principal component analysis on the iron blend where the average was obtained from averaging the first three and last three dates in our data set. The averaged spectrum over all dates is shown for reference at the top of the left-most figure.}
\label{fig:5169combo}
\end{figure*}

	In order to better understand the spectral lines that have been shown to have disk absorption during mid-eclipse \citep{Str14}, certain lines (see Table 2) were evaluated for changes over time. Some lines showed the formation of doublets around RJD 55399 such as the iron blend line at 5169.1{\AA} and 5169.3{\AA} (Figure~\ref{fig:5169combo}). The line is unshifted until just before first contact where a noticeable redshift begins and peaks around RJD 55200. On RJD 55399, clear doubling has begun in the line which persists and strengthens until after fourth contact. Right before third contact begins, the line begins to show blueshifting that persists all the way through fourth contact.
    
    Another line studied in detail was the Fe II line at 4629.33{\AA}, which traces the mass transfer stream the best. Figure~\ref{fig:4629combo} (see Appendix A) mostly shows line behavior during third contact as well as the out of eclipse behavior. As can be seen, line tripling persists through most of the third contact with the highest velocity occurring around RJD 55606. By the end of third contact the line is only doubled tracing background star and disk, but losing the contribution from the mass transfer stream. Also of note, is the extreme blueshift that begins to occur around mid-eclipse and peaks through third contact. 
    
    The iron line 5110.4{\AA} was also examined in detail because of the third-contact line strengthening observed. As can be seen in Figure~\ref{fig:5110combo}, the line is mostly non-existent before and after third-contact, but strengthens significantly during third contact showing strong blueshifted absorption.

In order to illustrate the significant differences over time we have explored differentials between spectra and averages of those spectra, using the principles of singular value decomposition (SVD) and/or principal component analysis (PCA) -- see \citet{Sku02} and \citet{Cas05}.

The principal component analysis was performed by first averaging the wavelengths and intensities of all epochs in our data set, for the case of 5169.3{\AA}, or by only averaging the first three and last three dates in our data sets as in the cases of 4629.33{\AA} and 5110.4{\AA}. Once an average line profile was obtained, a ratio was taken between the average line profile and each individual line profile in our range of dates. These results were plotted against radial velocities of each line with zero velocity being defined as the rest wavelength of the line. The results are shown alongside evolution plots in Figure~\ref{fig:5169combo} and Figures~\ref{fig:4629combo}-\ref{fig:5169c} in the appendix.

\subsection{Details of Stream Appearance}

\begin{figure*} 
\centering
\includegraphics[width=0.9\linewidth] {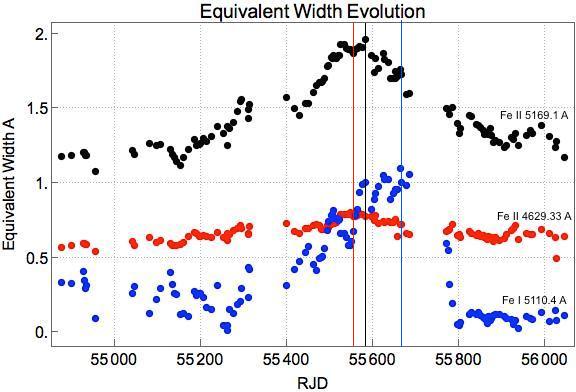}
\caption{This plot shows the equivalent width evolution for Fe II 4629.33{\AA} (middle/red line), Fe II 5169.1{\AA} (top/black line), and Fe I 5110.4{\AA} (bottom/blue line) together. Fe II 5169.1{\AA} and Fe II 4629.33{\AA} equivalent widths are measured in angstroms whereas the Fe I 5110.4{\AA} equivalent width values have been multiplied by 10 so that the maximum equivalent width date for each line can be seen. The vertical grid lines each correspond to the associated equivalent width maximum making it clear the distinction between the high-excitation potential lines and the low-excitation potential line. The gaps around 55000, 55400, and 55725 are due to sun blockage.  }
\label{fig:EWcombo}
\end{figure*}

The radial velocities over time for certain lines, such as Fe II 4629.33{\AA}, revealed the presence of a mass transfer stream that also shows up in the evolution plot of Fe I 5110.4{\AA}. The appearance during a fairly small window of time near third contact called for a more in-depth look of line evolution during RJD 55500-55800, which helped to reveal sub-structure within the disc. The three spectral lines studied for this investigation were the same as those presented in the previous section. The two Fe II lines have excitation potentials of approximately 2.8eV and, for the most part, trace the same features in the disc. The more interesting line is Fe I 5110.4{\AA}, which has an excitation potential of 0eV and is only prominent around third-contact. Extended evolution plots as well as complementary PCA plots were done for each line over RJD 55000-55800 (Figures~\ref{fig:4629a}-\ref{fig:5169c}) and equivalent widths were calculated for each line for every available date (Figure~\ref{fig:EWcombo}).

Originally, we just wanted to look closer at the dates right around third contact because of the presence of the mass-transfer stream in this window. However, by focusing in on RJD 55600-55700 we were alerted to features in all lines that left us with more questions, so we extended our evolution and PCA plots to include 100 days before RJD 55500 and 100 days after RJD 55800. Each line seemed to show slightly different dates of maximum velocity as well as maximum line strengths. In order to quantify these questions we decided to measure the equivalent widths for each line over the entire data set.

Measuring the equivalent widths of the high-excitation lines showed that both Fe II 4629.33{\AA} and Fe II 5169.1{\AA} have an overall rise and fall in line strength over the course of the eclipse. Each line reaches an absolute minimum at RJD 54956, shortly after data collection began, before steadily rising to a maximum equivalent width before third-contact begins. Fe II 4629{\AA} reaches its maximum equivalent width at RJD 55569, 13 days before Fe II 5169{\AA} reaches its maximum at RJD 55582.  These two lines share other common features including several shared local maxima and local minima. Each line reaches a local maximum on RJD 550823, which then falls to a local minimum on RJD 55143 for Fe II 4629{\AA} and RJD 55153 for Fe II 5169{\AA}. They also share another local minimum at RJD 55263 before rising to a maximum at RJD 55293 for Fe II 4629{\AA} and RJD 55296 for Fe II 5169{\AA}. Finally, there is another exact match on RJD 55429 where both lines experience a local minimum. Overall, these lines have the same overall shape and trend with very similar patterns of local maxima and minima that occur on almost the exact same dates with a few slight discrepancies. 

Another line, for which equivalent widths have only been calculated around third-contact (due to lack of signal outside this time-frame), is Fe I 5110.4{\AA}, a low excitation line that exhibits a pattern very different from the higher energy lines discussed above.  The equivalent width plot shown in Figure~\ref{fig:EWcombo} shows a minimum around 2nd contact (RJD 55250) before rising to the maximum equivalent width at RJD 55664, which is about 90 days after Fe II 4629{\AA} and Fe II 5169{\AA} reach maximum equivalent width. After this maximum, there is a rapid fall-off in equivalent width with values dropping by $\sim$95 percent over 112 days. After this drop off the values hover very close to zero for the remainder of the data set. 

\begin{figure*} 
\centering
\includegraphics[width=.5\textwidth]{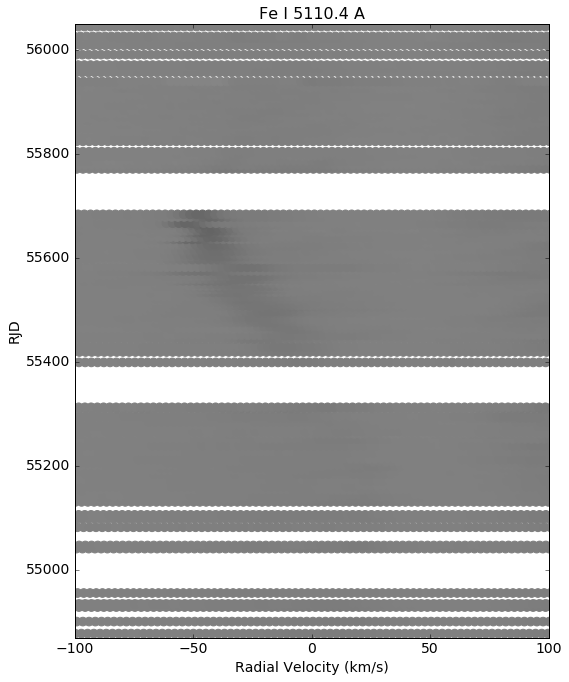}\hfill
\includegraphics[width=.5\textwidth]{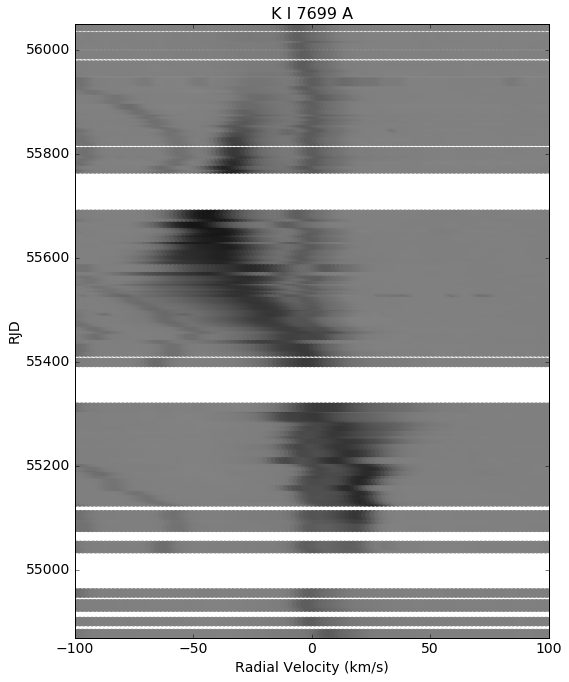}\hfill
\caption{Time-series intensity maps for two lines that show strong third-contact phenomena. The K 7699 line clearly shows the motion of the disk, while the Fe I 5110.4 line shows evidence of the mass-transfer stream. For a similar figure to the one on the right, refer to Strassmeier et al. (2014).  RJD timing is as presented in Table 1.}
\label{fig:tspec}
\end{figure*}


These timing differences between the roughly 3 eV excitation potential lines and the 0 eV line hints at a sub-structure in the disk characterized by the temperature gradient of the disk. Another line mentioned by \citet{St11} was He 10830{\AA} which had a maximum equivalent width of approximately 2{\AA} that occurred around the mid-eclipse at RJD 55400. Being a line of helium, the corresponding excitation potential is much higher at approximately 20 eV. For this reason the helium would need to be concentrated near the center of the disk to receive the highest energy photons from the companion star. At mid-eclipse the center of the disk is approximately lined up with the background F-star, which would explain the enhanced absorption for He 10830 at mid-eclipse. Next, we see lines of approximately 3 eV peaking roughly 50 days before the beginning of third-contact which reflects the movement of the disk away from the center of the background star. The helium line decreases in equivalent width as the Fe II lines begin to increase in equivalent width. Finally, around third-contact we see the lowest excitation potential line peak in equivalent width enhanced by the background F-star and presumably the mass transfer stream. The overall pattern we are seeing is a peak in equivalent width for the highest energy lines followed in succession by lower and lower excitation potential lines. Future work could focus on extending these findings to more lines and on determining a temperature function to describe the disk.


\subsection{TripleSpec Observations of He I 10830{\AA} Lines and Pa-$\beta$}
In addition to the optical spectra obtained by ARCES, described above, a co-mounted near-infrared spectrograph captured data on the same schedule (Table 1).  TripleSpec provides wavelength coverage from 0.95 to 2.46 microns with a spectral resolution of 3500 using the 1.1 arcsec slit.

\begin{figure*} 
\centering
\includegraphics[width=.45\textwidth]{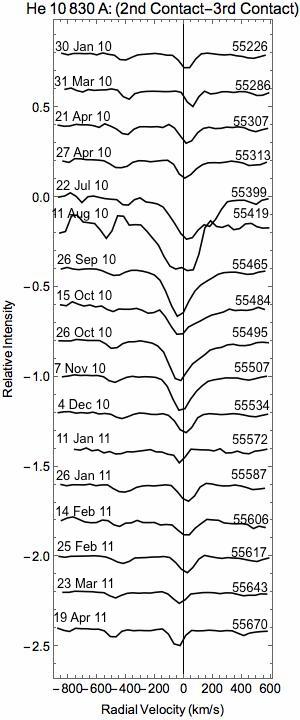}\hfill
\includegraphics[width=.454\textwidth]{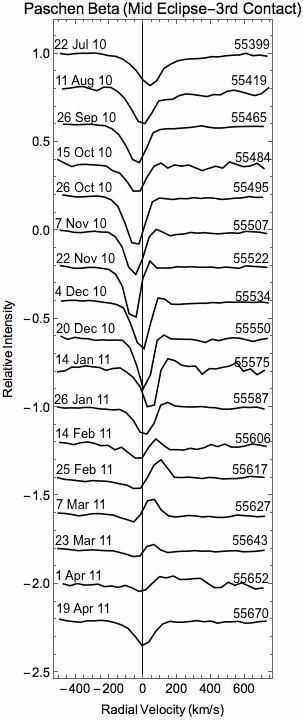}\hfill
\caption{Variation of the He I 10830{\AA} and  Pa-$\beta$ line profiles during the eclipse of $\epsilon$ Aurigae, as seen with the TripleSpec instrument.}
\label{fig:tspec}
\end{figure*}

Previously reported changes in the Helium I 10830{\AA} line \citep{St11} reveal a large increase in absorption equivalent width during mid-eclipse.  The more extensive coverage with TripleSpec confirms this behavior and enables study of details of the mid-eclipse, per Figure~\ref{fig:tspec}, including a suggestion of a local minimum near RJD 55475, roughly 1 AU past mid-eclipse. This offset could relate to the position of the disk inner wall.  We deduce temperatures and densities from this measurement in a later subsection.

Equally interesting was the appearance, starting after mid-eclipse and especially during third contact, of a P Cygni profile in the Paschen $\beta$ line at 1.28 microns. The blue-edge velocity is approximately --100 km s$^{-1}$, presumably arising from the energetics due to the stream-disk collision.

\section{Discussion}



We interpret the three types of velocity variations seen in Figure~\ref{fig:4629interp} as follows: the upper curve represents the F star orbital motion (amplitude a few km/sec), while the middle curve represents rotation of a disturbed portion of the disk (amplitude tens of km/sec), perhaps related to the extra feature near RJD 55600 - the impact of a mass transfer stream, briefly illuminated by the F star during third contact.  This third feature is the mass transfer stream, discovered by \citet{Gri13} but not revealed with this degree of detail, prior to this ARCES coverage.  Velocity curves presented by \citet{Str14} drew our attention to these high velocity features, but their temporal coverage was limited by instrument downtime, close to third contact.  We posit that the low and high eccentricity lines as defined by \citet{Str14} are defined by variation types 1 and 2, respectively, in our nomenclature.  Type 3 is the disrupted portion of the disk, caused by stream impact primarily among azimuths near third contact. 

The diversity of line variation types indicate structural elements of the disk system, differing by opacity and/or stream impact effects.  Low opacity, high excitation lines like Si II show no eclipse.  Hgiher opacity lines trace the disk bulk and its Keplerian rotation.  Other low excitation lines of Fe II and K I reveal the stream impact portion of the disturbed disk -- refer to line details in  Table 1. Strassmeier et al.reported that highest velocity disk lines show more eccentricity (their Fig.12). We interpret the increased velocity dispersion of the absorption-disk lines during egress as being due to a hot (or warm) spot on the trailing section of the disk -- consistent with stream/impact and general the ingress-egress asymmetry often reported in the literature for this star.

These sets of velocities define the energetics and must be shown to be consistent with a binary star mass ratio, once that gets determined.  Meanwhile, we consider the following estimates: adopting a 6 M$_{\odot}$ B star at the center of the disk, and an outer disk radius of 4AU as defined by the CHARA+MIRC interferometric observations, the Keplerian speed of the outer disk should be 36 km/sec, which is about that seen in type 2 variation.  The 60 km/sec represents a terminal velocity of material accelerated toward the B star from a large distance (e.g. the inner Lagrangian point), only $\sim$6AU distant, using these numbers.  The rate of deceleration, -60 to -40 km/sec over a span of $\sim$50 days, is a result of ballistic infall of material from the inner Lagrangian point and collision with transverse disk material. \\


\subsection{Evolutionary Models}

In order to better understand and make predictions about the masses and evolutionary future of epsilon Aurigae, a multi-parametric chi-squared minimization study of published evolutionary models was performed, similar to the method described by \citet{Men14}. We compared previously measured quantities, including the temperatures and luminosities of both components as proposed by \citet{Hoa10}, with those predicted by \citet{Van11} models of conservative and non-conservative binary star evolutionary models. We were able to seek models that best fit the observed data. Higher initial mass models are required for the F star to reach log luminosity of 4.5 or higher.  One interesting model identified in this way was 09strong, with a 20 day initial period.  This model features initial masses of 9 and 8.1 solar masses (q=1.11).  This model was evolved into a post-mass exchange binary, at step 1088, when t=32.825 Myr and masses have changed to 2.4 and 14.04 solar masses (q=0.17).  At that point, one component reached virtually the same luminosity as the originally more massive star (log L = 4.1, Log Te = 3.875). In a separate paper,  \citet{Gib18} use MESA binary models to further explore the evolutionary status of the epsilon Aurigae system. \\

\subsection{SHAPE Visualization of Disk and Stream}

SHAPE 5.0 is a morpho-kinematic modeling and reconstruction tool used for astrophysical objects (http://www.astrosen.unam.mx/shape/index.html). With SHAPE, users can input any known physical parameters of their objects and can augment single bodies with clouds, disks, and other features. It is also possible to specify chemical abundances as well as emission and absorption lines. Automatically created with each physical model is a velocity-space model, as well as a a graph of the object's velocity curve. Users also have the capability to change the frame through which the object is being viewed, by changing inclination and by rotating the object with respect to the observer. SHAPE is easy and quick to learn, making it an effective tool for modeling and interpreting astrophysical objects. One problem with SHAPE was the tendency for files to be unsaved or deleted in between opening and working with them. 

\begin{figure} 
\begin{center}
\includegraphics[width=1.0\linewidth] {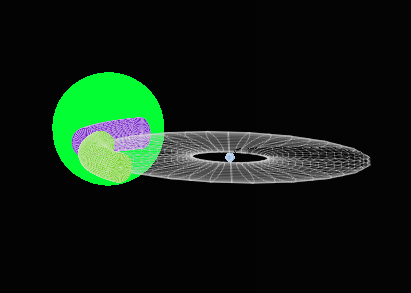}
\caption{SHAPE software visualization of the disk with the stream component representation at third contact projected against the background F-star (large green circle). The mass transfer stream is represented as the two curved cylinders on the left edge of the disk. The inclination of the system has been set at 80 degrees for the purpose of visual clarity.}
\label{fig:shape}
\end{center}
\end{figure}

\begin{figure*} 
\begin{center}
\includegraphics[width=0.8\linewidth] {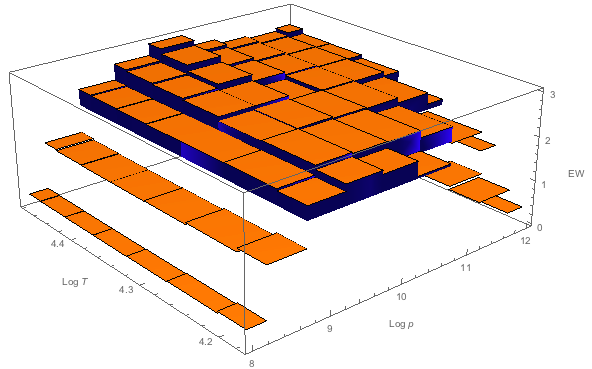}
\caption{A 3-Dimensional image showing how the equivalent width of He 10830{\AA} varies with density and temperature. Temperatures were run from 15000 K to 30000 K, with a step-size of 2500, and the log of the hydrogen density ranged from 8 to 12 cm$^{-3}$, with a step-size of 0.5. The bottom of the blue squares indicate the location of a 2.0{\AA} value for the equivalent width, which was the maximum observed value  for He 10830{\AA}.}
\label{fig:cloudyEW}
\end{center}
\end{figure*}

SHAPE was used to interpret and model the mass transfer stream observed during third contact. A small sphere was set in the center of the frame and surrounded by a disk, representing the B star and the accretion disk. The primary F star is set in the background moving past the accretion disk and there are two separate "clouds" representing the observed mass transfer stream (Figure~\ref{fig:shape}). The cloud farther out from the disk represents the higher velocity component that is illuminated by the F star first, and then transitions into the slower, closer in cloud as the eclipse ends. The deceleration from --60 km/s to --40 km/s over the projected distance of one F star radius is indicative of a 6 to 8 solar mass central star inside the disk, assuming the disk radius scale derived by \citet{Klo15} at the 737 pc distance.  The mass function implies an F star mass of 6 solar masses (q=0.78).  For the 1 kpc distance, the implied mass inside the disk is 11 solar masses, and for the F star, 12 solar masses (q = 1.1).

One challenging aspect of the stream velocity at third contact is why it only briefly appears between RJD 55600 and 55650, which is only 0.3 percent of the orbital phase.  This suggests the higher speed stream closely parallels the edge of the opaque disk.  Broadband polarization reported by \citet{Kem86} spikes after mid-eclipse and during third contact, which suggests strong forward scattering of light by circum-disk material during those phases.

\subsection{CLOUDY fitting of the He I 10830 $\AA$ line strength}

CLOUDY \citep{Fer98} is an internationally used spectral synthesis code designed to simulate conditions in plasmas ( http://www.nublado.org/ ), such as circumstellar and interstellar matter, exposed to a radiation field under a broad range of conditions. CLOUDY also has the ability to predict emission and absorption spectra, which can be compared with the observed data.  CLOUDY includes the 30 lightest elements in its calculations and the relative concentrations of these elements can be modified. The ionic and molecular emission data used in CLOUDY comes from the CHIANTI database \citep{Der97}, the LAMDA database, as well as its own atomic and molecular Database with level energies taken from NIST \citep{Tur16}. For our purposes, CLOUDY was used to interpret the equivalent widths of one of the lines that experienced broadening and strengthening during the eclipse: He 10830{\AA}. 

To generate equivalent widths from the CLOUDY code, certain parameters had to be specified in order to compute a continuum file that would contain the calculated intensities over our selected range of wavelengths. For the input parameters we varied log temperature and log density, to see their effect on the line strength, while other parameters such as geometry, radius, and chemical abundances were not varied. Temperature was run from 15000K to 30000K with a step size of 2500K, or in log terms, from 4.17 to 4.47. The log of the hydrogen density was varied from 8 to 12 with a step size of 0.5. For our code, we assumed an open geometry with an inner and outer radius set at 1AU and 5AU, respectively. Chemical abundances were assumed to be the same as solar abundances, and CLOUDY uses values provided by \citet{Gre10}. The code was iterated twice and the continuum results were exported to a text file. Equivalent width was plotted as a function of log density and log temperature, as can be seen in Figure~\ref{fig:cloudyEW}. The blue region shows the computed widths in excess of the observed maximum equivalent width of He I 10830{\AA}, 2{\AA} at mid-eclipse \citep{St11}.

Adopting a scenario of a disk with a central hole around a hot companion star, the He I 10830{\AA} naturally arises from that source, characterized by the CLOUDY results as fitted by log T = 4.1 to 4.2 and log n $\sim$ 11 (the falloff on the right hand side of the graph).  The density can be compared with results for the outer disk (third contact) obtained by analysis of infrared CO bands, yielding outer disk T$_{exc}$ = 1050 to 1275K and n$_{H}$ $\sim$ 3 x 10$^{10}$ cm$^{-3}$ \citep{St15}.  We conclude that the CLOUDY models support the scenario and provide constraints on density and temperature gradients in and around the disk.

\section{Conclusions}

Spectroscopic monitoring of $\epsilon$ Aurigae during its recent eclipse, using the Apache Point Observatory ARCES and TripleSpec instruments, has helped resolve detailed substructure in the disk and stream at selected phases.  In particular, the amplitude of the disk velocity is better defined, and the appearance and velocity of the stream is more well defined. Variations in the He I 10830{\AA} absorption line strength during mid-eclipse was shown to relate to the inner disk structure. The hydrogen Pa-$\beta$ line exhibits a P Cygni profile during third contact, suggestive of an expansion region related to a stream impact zone.  Having defined special phases around third contact, ideally more targeted observations can be arranged during the next eclipse.

One hope resulting from study of these data is another attempt to determine the distance to the epsilon Aurigae system - much debated in the literature. Recent work by \citet{Gui12} claimed to constrain distance estimates to 1.5 $\pm$ 0.5 kpc, using correlations between distance measurements to stars in the vicinity of epsilon Aurigae with interstellar absorption and reddening properties. However, that study rests on several assumptions, including uniformity of the interstellar medium (ISM), enforcing strictly linear correlations between distance-sensitive properties, and spectral type calibration uncertainties that factor into their {\it single star} MESA models and distance moduli. Distances based on different DIBs disagree with each other, so reliance on a single one also can be highly suspect \citep{Fri11}. Each of these assumptions contribute to a substantial cumulative uncertainty in their distance estimate. It is instructive to note that several binary stars exist in the  Aurigae field.  For example, B-type stars are fairly common within a few degrees, and indications are that the epsilon Aurigae primary could have arisen from a B-type main sequence star, and its disk may contain a B-type secondary star. Similarly, epsilon Aurigae has a moderate eccentricity e = 0.227 (Stefanik et al. 2010). Two neighbor stars, which also are \citet{Gui12} sample stars, turn out to be high eccentricity B-star binaries. These stars are HD 31894 (e=0.61, estimated stellar mass of 9M) and HD 31617 (e=0.76, estimated primary stellar mass of 9 to 15M). The formation of such binaries represents a challenge to star forming theories. Another neighbor, HD 30353 (A5Iap), was suggested to be a Case C mass transfer binary \citep{Lau70}. Collectively, these are clues to binary star evolution in this region of our galaxy.  We argue that this may be so, and will continue to explore the implications thereof in future work. The distance moduli for these B star binaries and angular proximity to epsilon Aurigae argue for a distance less than 1 kpc. After the first version of this paper was submitted, the GAIA Data Release 2 included a new parallax for epsilon Aurigae of 2.4 $\pm$ 0.5 milli-arcsec, corresponding to a distance estimate of 350 to 525 pc, strongly supporting the lower mass, small mass ratio (q $<$ 1) model \citep{Gib18}.

\section{Acknowledgements}
This work was supported in part by the bequest of William Herschel Womble in support of astronomy at the University of Denver.  J.G. thanks the University of Denver Undergraduate Research Center for a Partners in Scholarship (PinS) grant, and the Native American Scholarship fund that helped his efforts in this research.  An earlier version of this paper appeared as  https://arxiv.org/abs/1612.05287.  The authors are grateful to the Astronomical Research Consortium (ARC) that operates the Apache Point Observatory 3.5 meter telescopes and its ARCES and TripleSpec instruments that provided the bulk of the data reported in this paper.  We thank Matthew Muterspaugh (Tennessee State University) for access to late-eclipse time-series data generated by Joel Eaton with their Automated Spectroscopy Telescope (AST), review of which led us to re-examine the ARCES data during third-contact (see section 3.4: Details of Stream Appearance). We also want to acknowledge observers and APO program leaders including Alaina Bradley, Jack Dembicky, Suzanne Hawley, Joseph Huehnerhoff, Russet McMillan, Sarah Schmidt, Nicholas Ule, Donald G. York, George Wallerstein, the APO staff and others for their contributions to the ARCES and TripleSpec data collection process.  We appreciate helpful comments from the reviewer and editor.








\appendix

\section{Extra Figures}

The following figures contain intensity and PCA plots for our lines of interest covering a wider range of dates and with a higher sampling rate. This way the changes over time can be seen in more detail.  Note that several plots cover intervals in time for a specific line, e.g. RJD 55500 - 55600, 55600 - 55700, 55700 - 55800.

\begin{figure*} 
\centering
\includegraphics[width=.50\textwidth]{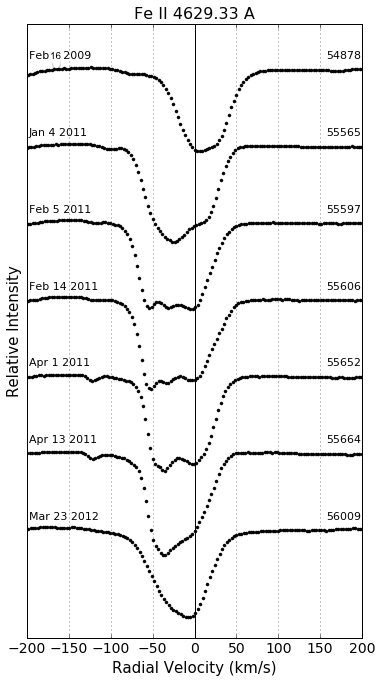}\hfill
\includegraphics[width=.50\textwidth]{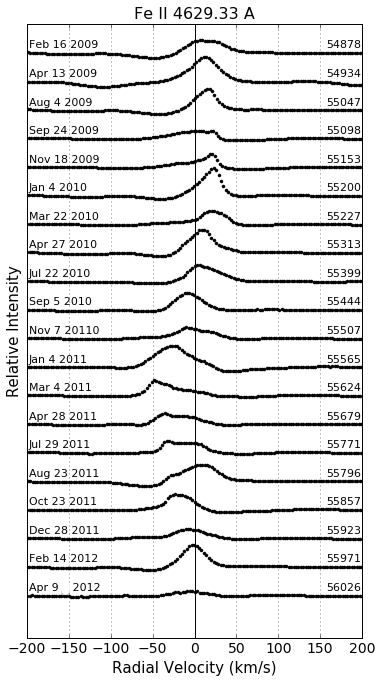}\hfill
\caption{This figure shows the evolution of the Fe II line at 4629.33{\AA}, during eclipse third-contact when the stream is most noticeable in the spectra. The left-most figure shows the intensity evolution around third contact clearly showing three minima that are representative of the background star, disk, and the mass transfer stream. Seven epochs were chosen to bring out details in the high-velocity, third contact material. The right-most figure shows the results of a first-order principal component analysis on 4629.33{\AA}, during the entire eclipse, where the average was obtained from the first three and last three spectra (out of eclipse phases) in our data set. For reference, the averaged spectrum is at the top of the left-most figure.  }
\label{fig:4629combo}
\end{figure*}

\begin{figure*} 
\centering
\includegraphics[width=.5\textwidth]{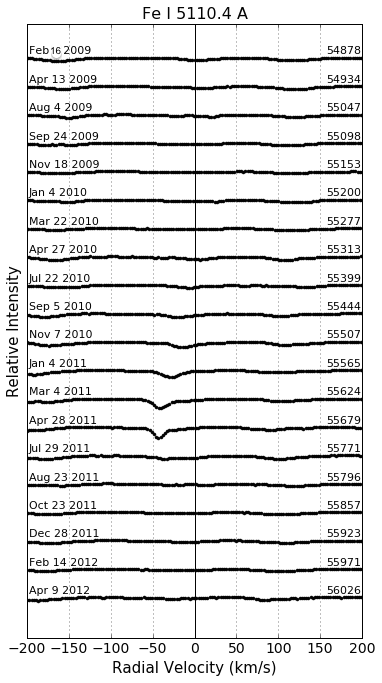}\hfill
\includegraphics[width=.5\textwidth]{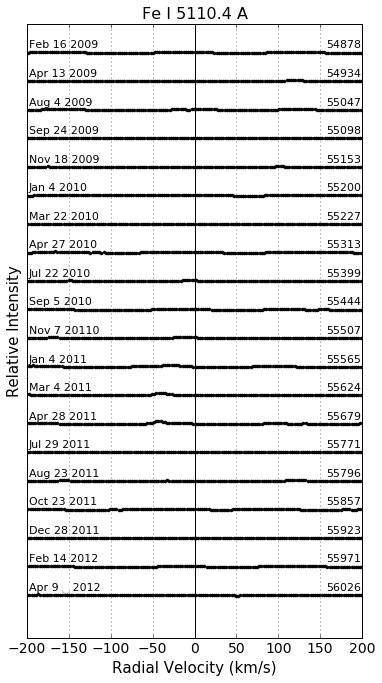}\hfill
\caption{Comparison between line evolution and a first-order ratio PCA for the line at 5110.4{\AA} similar to Figures 4 and A1. The average shown at the top of the left-most figure was obtained from averaging the first three and last three spectra in our data set.}
\label{fig:5110combo}
\end{figure*}

\begin{figure*} 
	\centering
	\includegraphics[width=.5\textwidth]{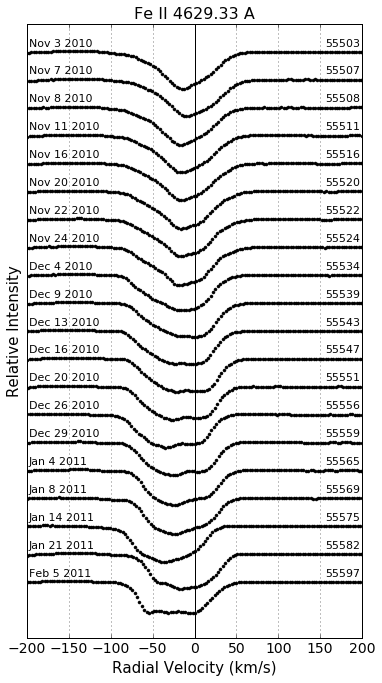}\hfill
	\includegraphics[width=.5\textwidth]{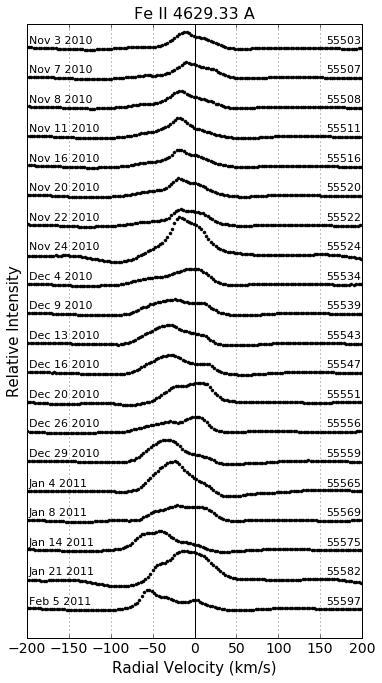}\hfill
	\caption{First of three epochs of coverage of the evolution and PCA plot for Fe II 4629.33{\AA} over the date range RJD 55500-55600, which spans post mid-eclipse up to the beginnings of third contact. Important to note in the evolution plot is the formation of doublet features (RJD 55551) and the triplet feature (RJD 55597).}
    \label{fig:4629a}
\end{figure*}

\begin{figure*}
	\centering
	\includegraphics[width=.5\textwidth]{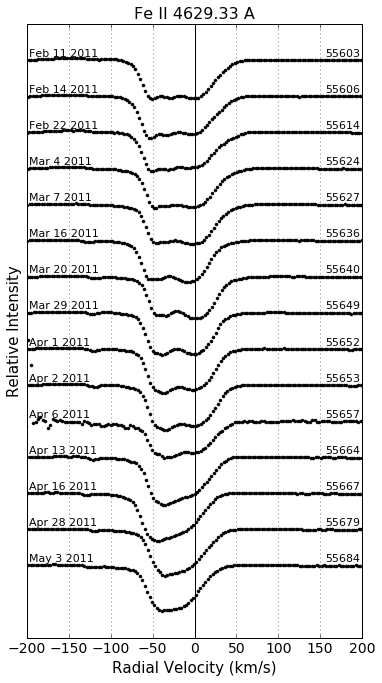}\hfill
	\includegraphics[width=.5\textwidth]{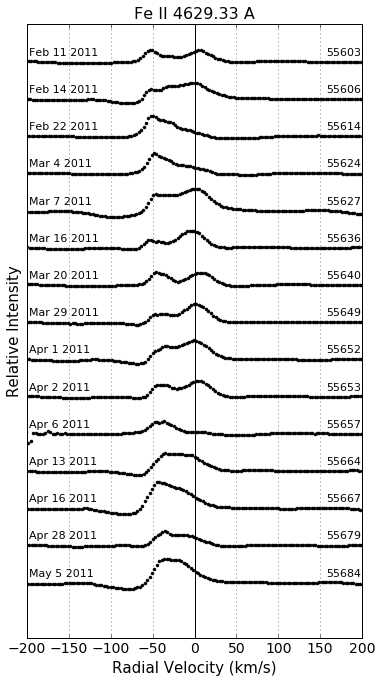}\hfill
	\caption{Second of three epochs of coverage of the evolution and PCA plot for Fe II 4629.33{\AA} over the date range RJD 55600-55700, which covers all of third-contact. Important to note is the changing line shape over the entire date range.}	
    \label{fig:4629b}
\end{figure*}

\begin{figure*}
	\centering
	\includegraphics[width=.5\textwidth]{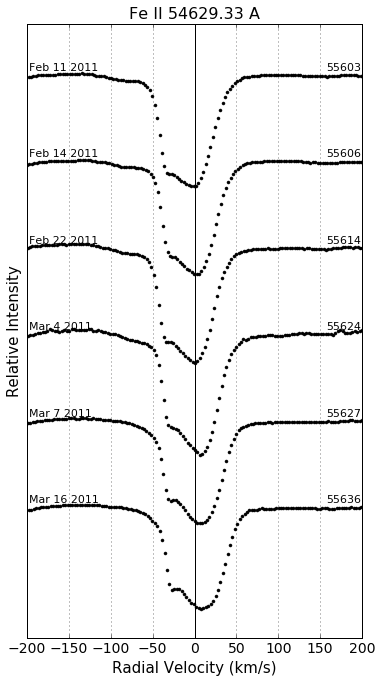}\hfill
	\includegraphics[width=.5\textwidth]{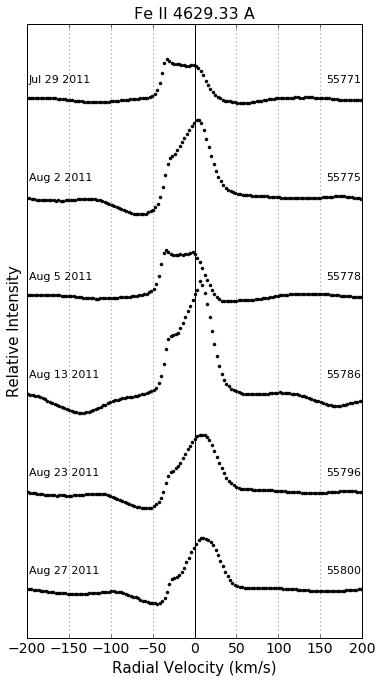}\hfill
	\caption{Third of three epochs of coverage of the evolution and PCA plot for Fe II 4629.33{\AA} over the date range RJD 55700-55800, which covers dates near fourth-contact.}	
    \label{fig:4629c}
\end{figure*}

\begin{figure*}
	\centering
	\includegraphics[width=.5\textwidth]{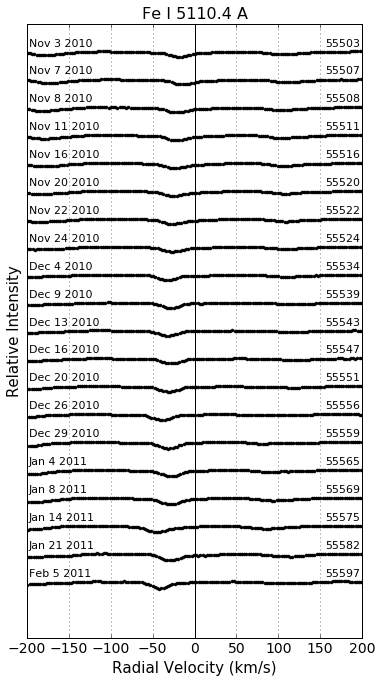}\hfill
	\includegraphics[width=.5\textwidth]{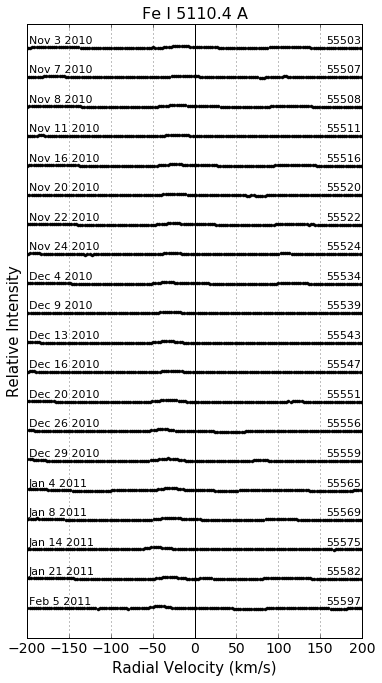}\hfill
	\caption{First of three epochs of coverage of the evolution and PCA plot for Fe I 5110.4{\AA} over the date range RJD 55500-55600, which spans post mid-eclipse up to the beginnings of third contact.}	
    \label{fig:5110a}
\end{figure*}

\begin{figure*}
	\centering
	\includegraphics[width=.5\textwidth]{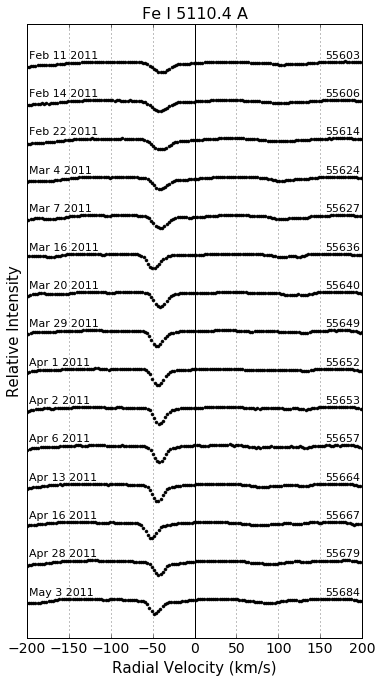}\hfill
	\includegraphics[width=.50\textwidth]{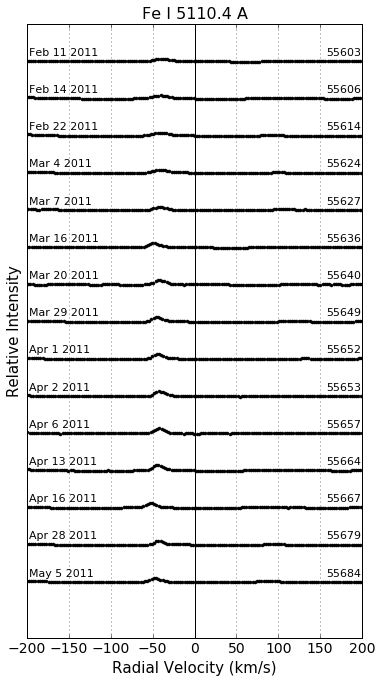}\hfill
	\caption{Second of three epochs of coverage of the evolution and PCA plot for Fe I 5110.4{\AA} over the date range RJD 55600-55700, which covers all of third-contact.}	
    \label{fig:5110b}
\end{figure*}

\begin{figure*}
	\centering
	\includegraphics[width=.5\textwidth]{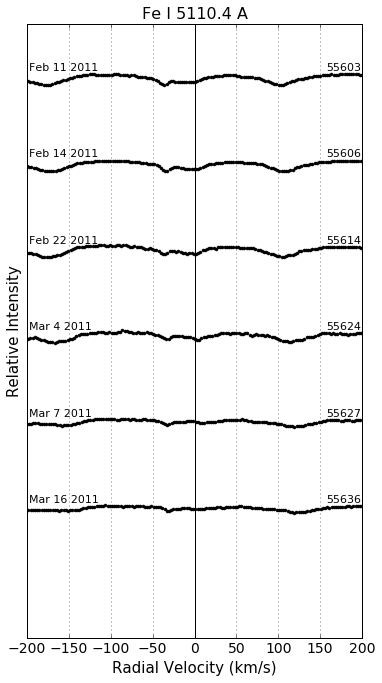}\hfill
	\includegraphics[width=.5\textwidth]{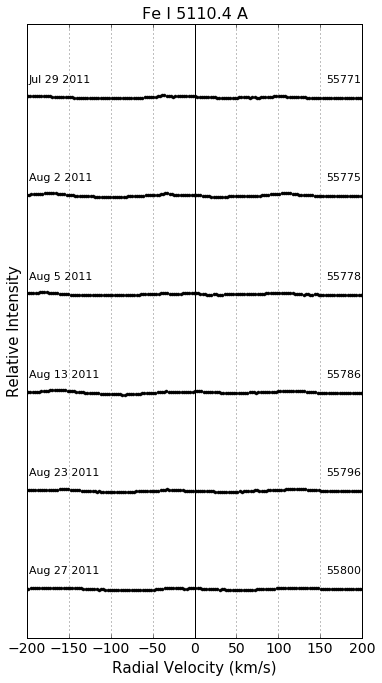}\hfill
	\caption{Third of three epochs of coverage of the evolution and PCA plot for Fe I 5110.4{\AA} over the date range RJD 55700-55800, which covers dates near fourth-contact.}	
    \label{fig:5110c}
\end{figure*}

\begin{figure*}
	\centering
	\includegraphics[width=.5\textwidth]{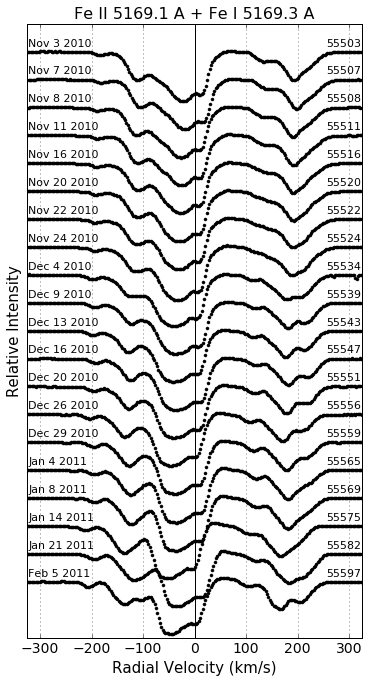}\hfill
	\includegraphics[width=.5\textwidth]{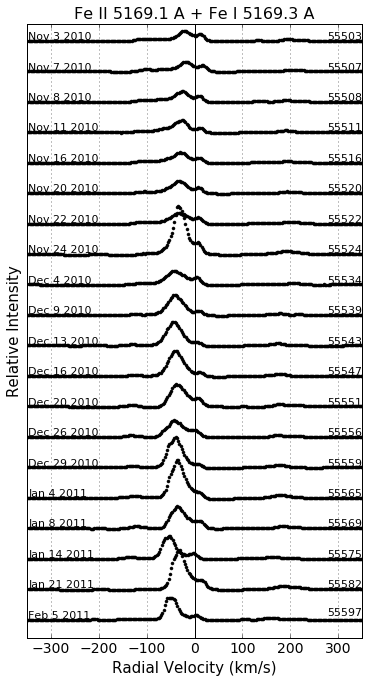}\hfill
	\caption{First of three epochs of coverage of the evolution and PCA plot for Fe II 5169.1{\AA} over the date range RJD 55500-55600, which spans post mid-eclipse up to the beginnings of third contact.}	
    \label{fig:5169a}
\end{figure*}

\begin{figure*}
	\centering
	\includegraphics[width=.5\textwidth]{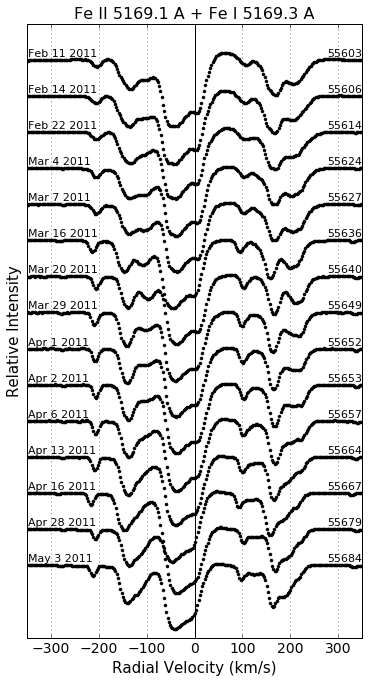}\hfill
	\includegraphics[width=.5\textwidth]{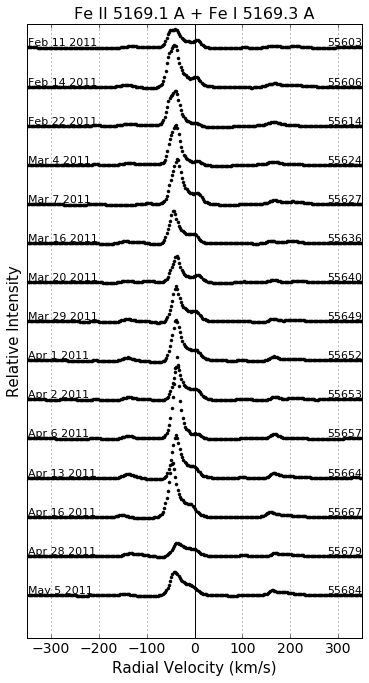}\hfill
	\caption{Second of three epochs of coverage of the evolution and PCA plot for Fe II 5169.1{\AA} over the date range RJD 55600-55700, which covers all of third-contact.}	
    \label{fig:5169b}
\end{figure*}

\begin{figure*}
	\centering
	\includegraphics[width=.5\textwidth]{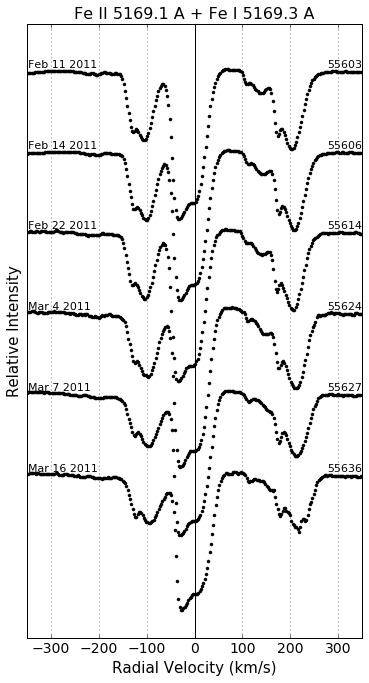}\hfill
	\includegraphics[width=.5\textwidth]{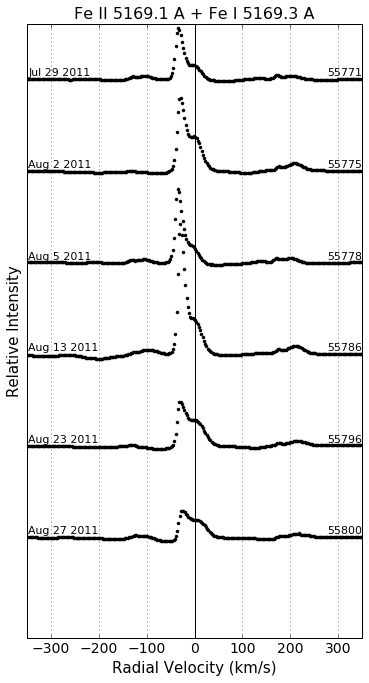}\hfill
	\caption{Third of three epochs of coverage of the evolution and PCA plot for Fe II 5169.1{\AA} over the date range RJD 55700-55800, which covers dates near fourth-contact.}	
    \label{fig:5169c}
\end{figure*}


\bsp	
\label{lastpage}
   \end{document}